# AlGaAs-On-Insulator Nonlinear Photonics

**Minhao Pu*, Luisa Ottaviano, Elizaveta Semenova, and Kresten Yvind****

*DTU Fotonik, Department of Photonics Engineering, Technical University of Denmark, Building 343, DK-2800 Lyngby, Denmark*

*mipu@fotonik.dtu.dk; **kryv@fotonik.dtu.dk

**The combination of nonlinear and integrated photonics has recently seen a surge with Kerr frequency comb generation in micro-resonators[1] as the most significant achievement. Efficient nonlinear photonic chips have myriad applications including high speed optical signal processing[2,3], on-chip multi-wavelength lasers[4,5], metrology[6], molecular spectroscopy[7], and quantum information science[8]. Aluminium gallium arsenide (AlGaAs) exhibits very high material nonlinearity and low nonlinear loss[9,10] when operated below half its bandgap energy. However, difficulties in device processing and low device effective nonlinearity made Kerr frequency comb generation elusive. Here, we demonstrate AlGaAs-on-insulator as a nonlinear platform at telecom wavelengths. Using newly developed fabrication processes, we show high-quality-factor ($Q$>$10^5$) micro-resonators with integrated bus waveguides in a planar circuit where optical parametric oscillation is achieved with a record low threshold power of 3 mW and a frequency comb spanning 350 nm is obtained. Our demonstration shows the huge potential of the AlGaAs-on-insulator platform in integrated nonlinear photonics.**

Desirable material properties for all-optical $\chi^{(3)}$ nonlinear chips are a high Kerr nonlinearity and low linear and nonlinear losses to enable high four-wave mixing (FWM) efficiency and parametric gain. Many material platforms have been investigated showing the different trade-offs between nonlinearity and losses[2-5,11-16]. In addition to good intrinsic material properties, the ability to perform accurate and high yield processing is desirable in order to integrate additional functionalities on the same chip and also to decrease linear losses. Silicon-on-insulator is the model system for integration[2]. It supports a large index contrast and shallow etch depths, enabling patterning sub-micron structures with smooth sidewalls. However, due to two-photon absorption (TPA) at telecom wavelengths, e.g. 1550 nm, high parametric gain[15] and Kerr frequency comb generation[16] have only been achieved by pumping at wavelengths beyond 2000 nm.

Aluminium gallium arsenide ($Al_xGa_{1-x}As$) was early identified as a promising candidate and even nominated as the "silicon of nonlinear optics"[9] when operated just below half its bandgap energy. It offers a high refractive index ($n$≈3.3) and a nonlinear index ($n_2$)[17-19] on the order of $10^{-17}$ W/m$^2$, a large transparency window (from near- to mid-infrared), and the ability to engineer the material bandgap by varying the alloy composition ($x$)[19]. Its bandgap can be tailored in such a way that TPA, which is the main detrimental effect for the FWM process, is mitigated and at the same time, the three-photon absorption is low[10] while a high material nonlinearity is maintained. Over the past two decades, efforts have been made to realize efficient nonlinear processes in AlGaAs waveguides[17-19]. However, the fabrication of such waveguides with very high and narrow mesa structures becomes very challenging and prevents advanced designs that go beyond simple straight waveguides. In addition, the low vertical index contrast of such waveguides limits the effective nonlinearity.

To enhance light confinement and relax the etching process requirements, we propose an AlGaAs-on-insulator (AlGaAsOI) platform as shown in Fig. 1a. In this layout, a thin $Al_xGa_{1-x}As$ layer on top of a low index insulator layer resides on a semiconductor substrate. Wafer bonding and substrate removal are used to realize the structure. In this letter, the aluminium fraction ($x$) is 17%, which makes the material bandgap 1.63 eV and the refractive index 3.33. Thanks to the large index contrast (~55%) of this layout, light can be confined in a sub-micron waveguide core. As the nonlinear parameter ($\gamma$) is highly dependent on the waveguide effective mode area ($A_{eff}$) as expressed[2] by $\gamma=2\pi n_2/\lambda A_{eff}$, an ultra-high effective nonlinearity of about 660 $W^{-1}m^{-1}$, which is orders of magnitude higher than that of a typical $Si_3N_4$ waveguide[5], can be obtained for an AlGaAsOI waveguide using a cross-section dimension of 320 nm×630 nm (see Methods). In addition, the waveguide dispersion dominates over material dispersion for sub-wavelength sized waveguides and therefore the group velocity dispersion (GVD) can be engineered from the normal dispersion of bulk material to anomalous dispersion (Fig. 1b), which is required to achieve parametric gain in nonlinear processes[15]. High index-contrast waveguide device performance is typically limited by linear loss induced by light scattering from surface roughness[16]. High quality epitaxial material growth and substrate removal is required to ensure smoothness for top and bottom waveguide surfaces while good lithography and dry etching processes define the waveguide sidewall surface roughness. A critical element is the choice of etch stop layer in the substrate removal process (see Methods) as e.g. hydrofluoric acid etching of Al(Ga)As leaves a too rough surface[20]. Fig. 1c shows a scanning electron microscope (SEM) picture of a fabricated AlGaAsOI waveguide with about 1.4 dB/cm linear loss. Combining the ultra-high effective nonlinearity and low loss, AlGaAsOI can be used for ultra-efficient nonlinear parametric processes such as low-threshold Kerr frequency comb generation once a high quality factor ($Q$) micro-resonator is realized.

Fig. 2a shows a SEM picture of an 810-μm long race-track-shaped AlGaAsOI micro-resonator. Fig. 2b shows a coupling gap of 170 nm for the resonator where the light propagating in the 450 nm-wide bus waveguide can be evanescently coupled to the 630 nm-wide curved (17.5 μm radius) waveguide of the resonator. The resonator works in the under-coupled regime and its transmission is shown in Fig. 2c for the transverse electric (TE) mode. Only one mode family with a free spectral range (FSR) of ~0.82 nm (98 GHz) is observed in the spectrum, which implies that the resonator waveguide with anomalous dispersion can be operated in a single-mode state. Therefore, the dispersion distortion induced by inter-mode interaction between different mode families can be completely avoided, which is preferable for Kerr frequency comb generation (especially temporal soliton formation) but not attainable in all nonlinear material platforms[21]. Fig. 2d shows the measured transmission spectrum for the resonance at 1589.64 nm and the measured linewidth is around 9.6 pm which corresponds to a $Q$ of ~165,600. The measured $Q$ for all the devices ranges from $1.5\times10^5$ to $2.0\times10^5$, which is more than an order of magnitude higher than previously demonstrated $Q$ for AlGaAs micro-ring resonators[22].

Kerr frequency comb generation is based on optical parametric oscillation (OPO), which relies on a combination of parametric amplification and oscillation as a result of the nonlinear FWM processes within the micro-resonator. As a continuous wave pump light is tuned into a cavity resonance to achieve thermal soft-locking[23], the built-up intra-cavity power triggers OPO at a critical power threshold when the round-trip parametric gain exceeds the round-trip loss of the resonator. To satisfy the momentum conservation in the FWM process, the pump energy can only be transferred to equispaced frequencies within the supported resonances of the micro-resonator[1] and thus form a frequency comb at output as illustrated in Fig. 3a. The measured spectrum of an AlGaAsOI micro-resonator is shown in Fig. 3b when a pump power of 72 mW was coupled into the bus waveguide. A frequency comb with the native line spacing (single FSR) spanning over about 350 nm was observed.

The threshold power in the bus waveguide can be estimated by the expression[24] $P_{th} \approx 1.54(\frac{\pi}{2})\frac{Q_C}{2Q_L}\frac{n^2 L A_{eff}}{n_2 \lambda_P}\cdot\frac{1}{Q_L^2}$, where $L$ and $\lambda_p$, are the cavity length and pump wavelength, and $Q_C$ and $Q_L$ are the coupling and loaded quality factors of the resonator, respectively. As $Q_L$ and $L$ are correlated, we measured the threshold power for micro-resonators with different cavity lengths (FSR ranging from 98 GHz to 995 GHz) to find their influences on threshold power. The obtained minimum threshold power was 3 mW for a ring-shaped micro-resonator (see inset of Fig. 4a) operated in the under-coupling condition with a $Q_L$ of about $10^5$. The output power of the primary OPO sideband increases significantly at threshold, as shown in Fig. 4a. Fig. 4b shows the measured output spectrum for this resonator with a pump power (4.5 mW) slightly above the threshold. It shows a typical OPO initial state in which widely spaced (multi-FSR spacing) primary sidebands are generated[25]. The measured threshold power for different devices as a function of $Q_L$ is shown in Fig. 4c, where most of micro-resonators are seen to have milliwatt-level thresholds, although micro-resonators with smaller FSR exhibit slightly higher threshold. The measured data follows the theoretically predicted threshold trend as shown by the coloured bands, and a sub-mW threshold power can be expected with further improvement of resonator design[26] and fabrication processes.

As the dynamics of Kerr frequency comb generation have been extensively studied[23-25] and mode-locked combs have been demonstrated[21,27], Kerr frequency comb technologies are approaching practical applications. Planar integration platforms are critical for practical systems because of their robustness and potential for a fully integrated comb system with on-chip light sources. Low threshold is crucial for the realization of such a system. Table 1 summarizes several planar integrated nonlinear material platforms where a frequency comb has been demonstrated at telecom wavelengths, including Hydex[4], $Si_3N_4$[5], AlN[11], Diamond[12] and AlGaAs. The linear refractive index of AlGaAs is the highest among these platforms, which makes it the most suitable platform for compact circuits. In addition, the material nonlinear index of AlGaAs is orders of magnitude larger than those of the other platforms. These intrinsic material properties make AlGaAsOI an ultra-efficient platform for nonlinear parametric processes. Therefore, even though the $Q_L$ of our device is relatively low, the OPO threshold power for the AlGaAsOI platform is the lowest compared with the other platforms. In line with the fast-growing hybrid integration trend to combine different materials in multiple levels on a single CMOS compatible chip, the AlGaAsOI platform is very promising for realizing a fully-integrated comb system.

Because of the wide transparency window of AlGaAs materials, the frequency comb can potentially be extended into the mid-infrared with proper dispersion engineering. Besides comb generation, we have recently demonstrated wavelength conversion of a serial data signal at a record high speed (beyond terabaud) in this platform[28]. An ultra-broad bandwidth and ultra-high efficiency make AlGaAsOI an excellent platform for optical signal processing in telecommunications. Moreover, AlGaAs exhibits strong $\chi^{(2)}$ effects[29] due to its non-centrosymmetric crystal structure. Therefore, the AlGaAsOI platform is also suitable for combining both $\chi^{(2)}$ and $\chi^{(3)}$ effects to obtain, for example, multi-octave spanning combs[30].

## Methods

**Device fabrication.** The preparation of AlGaAsOI samples includes wafer growth, wafer bonding, and substrate removal. A 320-nm thick layer of $Al_{0.17}Ga_{0.83}As$ was grown in a low-pressure metalorganic vapour phase epitaxy (MOVPE) reactor on a 50-mm GaAs (100) substrate with sacrificial layers. After depositing a 3-µm thick silica layer on the AlGaAs layer using plasma-enhanced chemical vapour deposition (PECVD), a 90-nm thick Benzocyclobutene (BCB) layer was used as a bonding polymer between the grown wafer and another 50-mm semiconductor substrate covered with a 10-nm thick silica layer. The bonding process was performed under partial vacuum (~$3\times10^{-2}$ Pa) at 250 $^{0}$C for one hour in a wafer bonding system, while a force of 750 N was applied to the wafers. Subsequently, the GaAs substrate and the sacrificial layers were removed by wet etching. It is crucial to use proper etch stop layer and etchant as the etching by-products may result in roughness and absorption sites on the AlGaAs film surface (Supplementary Fig. 1). Electron beam lithography (EBL, JEOL JBX-9500FS) was used to define the device pattern in the electron beam resist hydrogen silsesquioxane (HSQ, Dow Corning FOX-15). To get a smooth pattern definition for bent waveguides in micro-resonators, a multi-pass exposure was applied to mitigate the stitching between pattern segments fractured during the EBL (Supplementary Fig. 2). The device pattern was then transferred into the AlGaAs layer using a boron trichloride ($BCl_3$)-based dry etching process in an inductive coupled plasma reactive ion etching (ICP-RIE) machine. As the refractive index of HSQ is relatively low (similar to $SiO_2$), it was kept on top of the AlGaAs device pattern. Finally, clad in a 3-µm thick silica layer using PECVD, the chip was cleaved to form the input and output facets where nano-tapers enabled efficient chip-to-fibre coupling for characterization.

**Modelling.** Waveguide dispersion was calculated using a finite-element method mode solver (COMSOL) for the TE mode in the wavelength range between 1300 nm and 1800 nm. The Sellmeier equation was used to incorporate the dispersion of $SiO_2$ and the modified single effective oscillator model was used for AlGaAs. The dispersion of HSQ, a leftover electron beam resist on top of AlGaAs, was assumed to be the same as that of $SiO_2$.

**Measurement of nonlinearity.** A FWM measurement was performed using two continuous wave signals (one for the pump and one for the signal) in a 3-mm long AlGaAsOI nano-waveguide with a cross-section of 320 nm×630 nm. The pump-to-signal detuning was less than 1 nm to ensure phase matching[14]. We measured the dependence of conversion efficiency $\eta$ on pump power (Supplementary Fig. 4) and extracted the nonlinear coefficient $\gamma$ by fitting the expression[18]

$$\eta = \frac{P_{idler}(L)}{P_{signal}(L)} = \left(\gamma P_{pump}(0) L_{eff}\right)^2$$

where $L_{eff} = (1 - e^{-\alpha L})/\alpha$ is the effective waveguide length and α the linear loss coefficient. The conversion efficiency was determined from the spectrum at the chip output. The pump power was estimated by subtracting the chip-to-fibre coupling loss from the fibre output power. The waveguide propagation loss (1.4 dB/cm) was measured at 1550 nm using the modified cutback method (Supplementary Fig. 3). Assuming the input and output coupling are the same, we estimated the coupling loss to be ~2.7 dB/facet by subtracting the waveguide loss from the fibre-to-fibre insertion loss (~6.3 dB). We obtained an ultra-high nonlinear coefficient: ~660 $W^{-1}m^{-1}$.

**Acknowledgements**

The authors would like to thank Jan W. Thomsen at Copenhagen University, Christophe Peucheret at Université de Rennes 1, Molly Piels, Hao Hu, Leif K. Oxenløwe and Jesper Mørk at the Technical University of Denmark for motivating and helpful discussions and comments on this manuscript. The authors acknowledge financial support from Villum Fonden via the Centre of Excellence Nanophotonics for terabit optical communication (NATEC). M. P. also acknowledges the Danish Research Council via the SiMOF project (11-117031).

**Author contributions**

M.P. designed the device. M. P., L.O., and E.S. fabricated the device. M.P. performed the device characterization and analysed the data. K.Y. conceived the idea and supervised the work. M.P. and K.Y. wrote the manuscript.

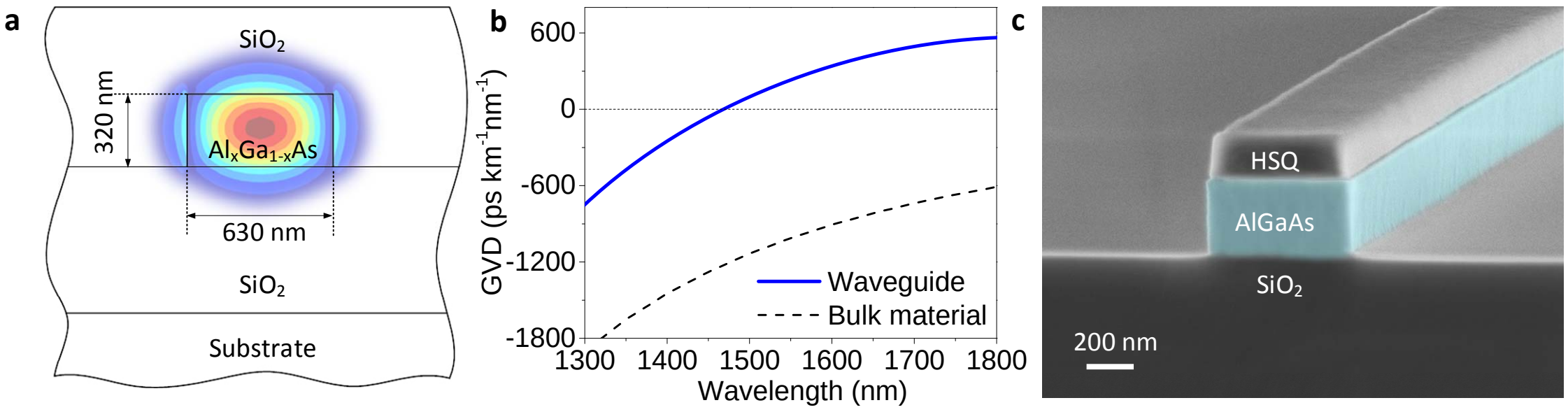

**Fig. 1 | AlGaAs-on-insulator nano-waveguide. a**, Schematic drawing of an AlGaAs-on-insulator (AlGaAsOI) nano-waveguide with the simulated electric field distribution for the fundamental transverse electric (TE) mode. **b,** Calculated group velocity dispersion (GVD) versus wavelength for the waveguide with 320 nm×630 nm cross-section dimensions. The dashed curve shows the GVD for bulk AlGaAs material. **c**, Scanning electron microscope (SEM) picture of an AlGaAsOI nano-waveguide (highlighted in blue) after the dry-etching process.

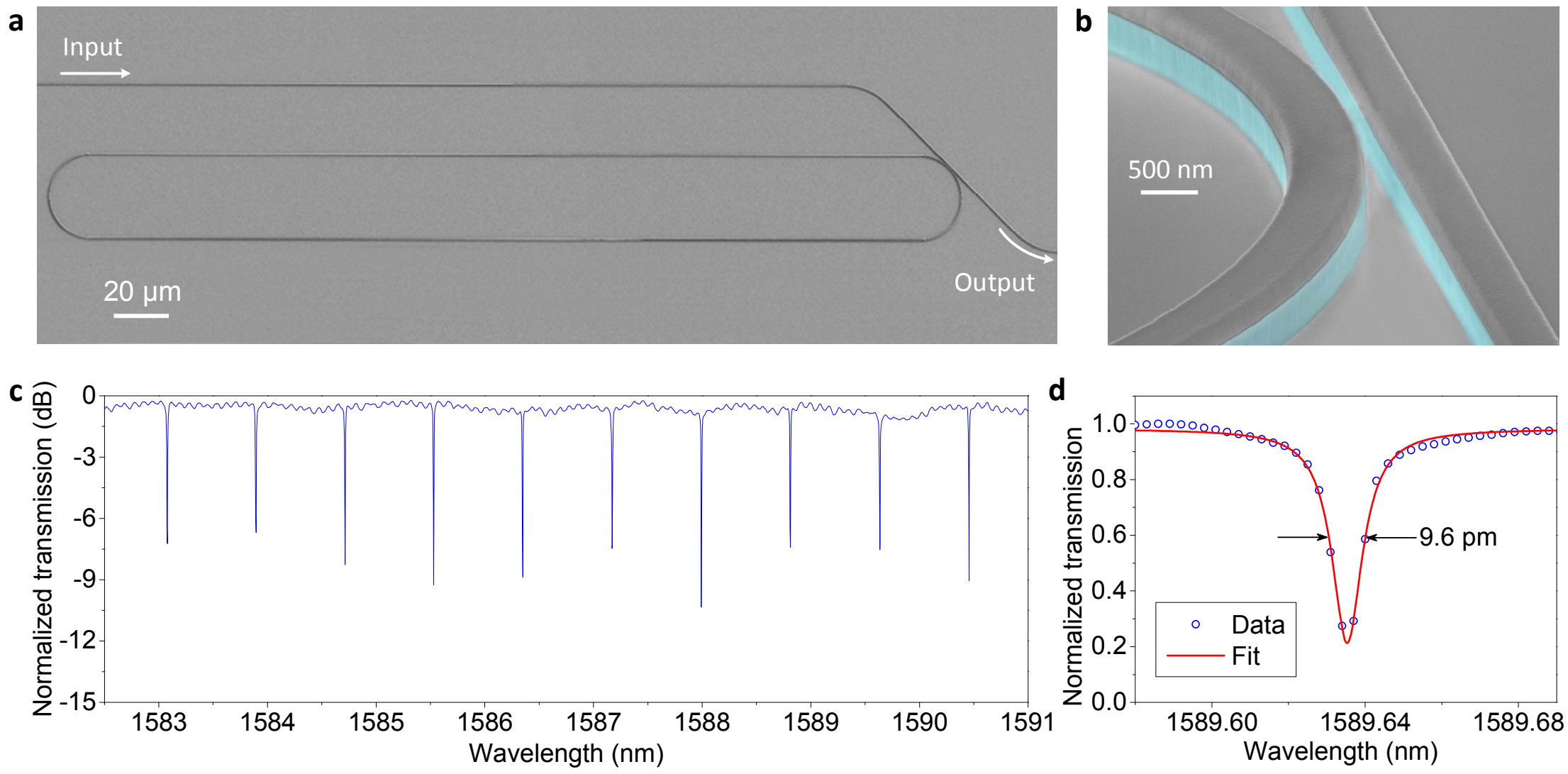

**Fig. 2 | AlGaAs-on-insulator microresonator**. **a,** Top-view SEM image of the AlGaAsOI micro-resonator showing ring and bus waveguide. **b**, Isometric view SEM image of the coupling region of the AlGaAsOI micro-resonator where the AlGaAs material is indicated in blue. **c**, Measured (normalized) transmission spectrum of the AlGaAsOI micro-resonator. **d**, Measured (normalized) transmission spectrum around the resonance of the AlGaAsOI micro-resonator at 1590 nm with a loaded quality-factor (*Q*) of ~165,600.

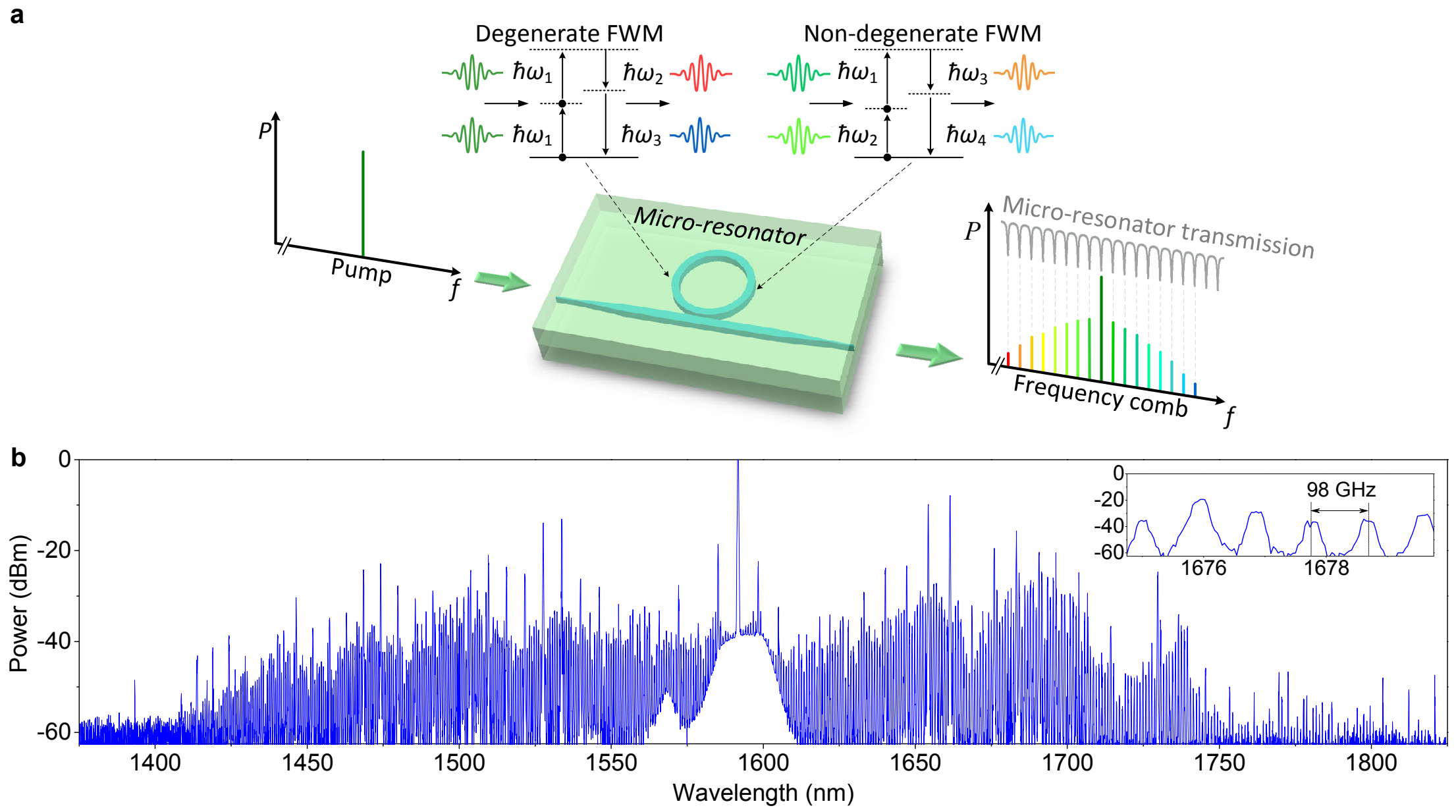

**Fig. 3 | Optical frequency comb generation in an AlGaAs-on-insulator micro-resonator. a**, A single-frequency pump light coupled into a high $Q$ micro-resonator enables efficient parametric processes (degenerate and non-degenerate four-wave mixing) in the resonator and allows new frequency generation in the supported modes of the micro-resonator. **b**, The output spectrum of an 810-µm long AlGaAsOI micro-resonator when a 72-mW pump was coupled into the resonator at ~1590 nm. A frequency comb was generated with a line spacing of 98 GHz over a wavelength range of about 350 nm.

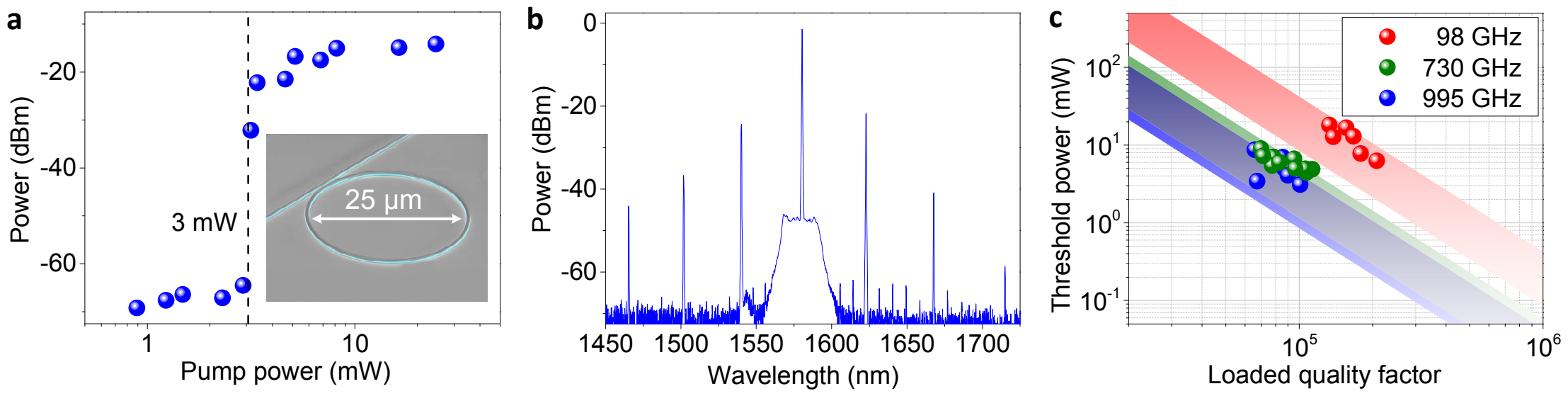

**Fig. 4 | Optical parametric oscillation threshold measurement**. **a**, Measured primary sideband power at the chip output as a function of pump power in the bus waveguide for a 25-μm diameter under-coupled micro-resonator with a FSR of 995 GHz as shown in the inset. A significant power increase was observed at the threshold of 3.1 mW. **b**, The output spectrum of optical parametric oscillation when the pump power was increased to just above the threshold power (4.5 mW) for the same micro-resonator in **a**. **c**, The measured threshold power for micro-resonators with three different FSRs: 98 GHz (red), 730 GHz (green), and 995 GHz (blue). The coloured bands shows the theoretical threshold power ranges for the different micro-resonators (from critical-coupling with $Q_C$=2$Q_L$ to under-coupling with $Q_C$=10$Q_L$).

**Table 1 Comparison of planar nonlinear platforms for frequency comb generation at telecom wavelengths**

| Material Platform | $n$ | $n_2$ (W/m$^2$) | Height (µm) | Width (µm) | Comb line spacing (GHz) | $Q_L$ | $P_{th}$ (mW) |
|---|---|---|---|---|---|---|---|
| Hydex[4] | 1.7 | $1.2\times10^{-19}$ | 1.45 | 1.50 | 200 | $1\times10^{6}$ | 50 |
| $Si_3N_4$[5] | 2.0 | $2.5\times10^{-19}$ | 0.71 | 1.70* | 580 | $2\times10^{5}$ | 50 |
| AlN[11] | 2.1 | $2.3\times10^{-19}$ | 0.65 | 3.50 | 435 | $8\times10^{5}$ | 200 |
| Diamond[12] | 2.4 | $8.2\times10^{-20}$ | 0.95 | 0.85 | 925 | $1\times10^{6}$ | 20 |
| $Al_{0.17}Ga_{0.83}As$ (this work) | 3.3 | $2.6\times10^{-17}$ | 0.32 | 0.62 | 995 | $1\times10^{5}$ | 3 |

*Taken as a base width with 20° sidewall angle.

# Supplementary information for

# AlGaAs-on-Insulator Nonlinear Photonics

**Minhao Pu*, Luisa Ottaviano, Elizaveta Semenova, and Kresten Yvind****

*DTU Fotonik, Department of Photonics Engineering, Technical University of Denmark, Building 343, DK-2800 Lyngby, Denmark*

*mipu@fotonik.dtu.dk; **kryv@fotonik.dtu.dk

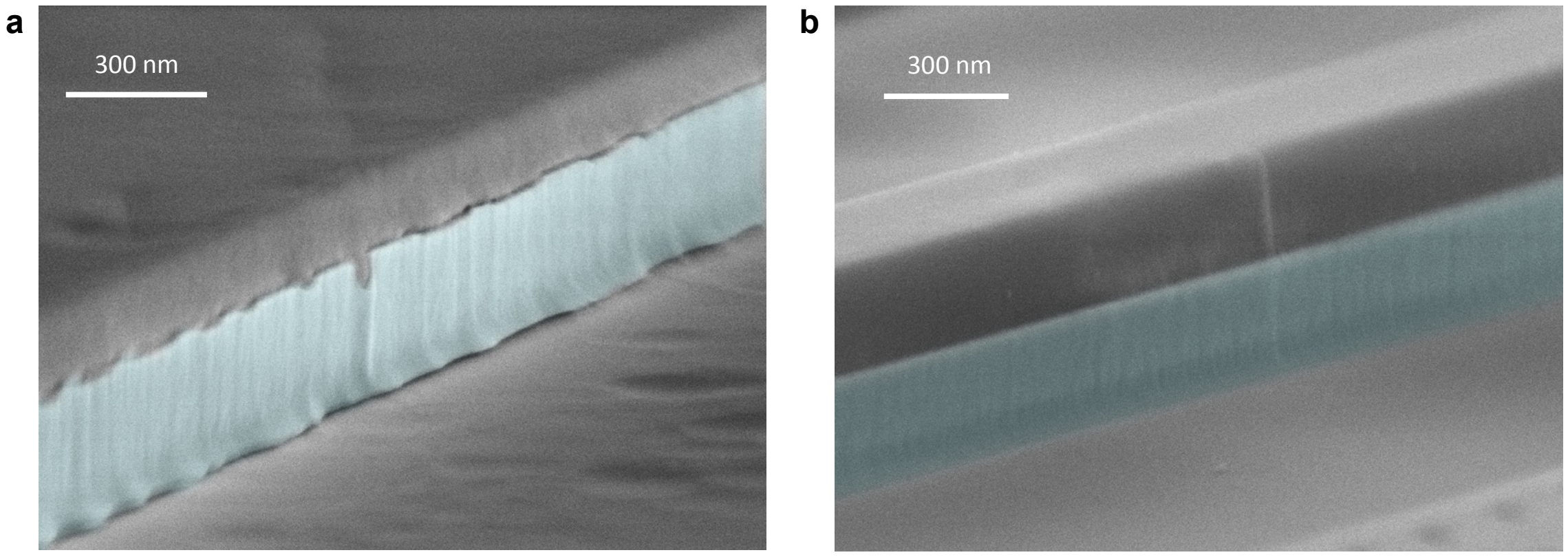

**Fig. 1 | Device surface roughness comparison.** Scanning electron microscope (SEM) pictures of etched patterns on AlGaAs-on-insulator (AlGaAsOI) wafers fabricated by using different substrate removal processes: **a,** Large roughnesses are observed at the top and bottom surfaces of the AlGaAs film when a conventional etch stop layer[1] is used. **b,** Significant improvement concerning surface smoothness can be observed when a proper etch stop layer was used[2]. AlGaAs material is denoted by the artificial blue colour in both pictures. Etch mask hydrogen silsesquioxane (HSQ) is left on top of AlGaAs patterns. The same dry etching processes are used in both cases.

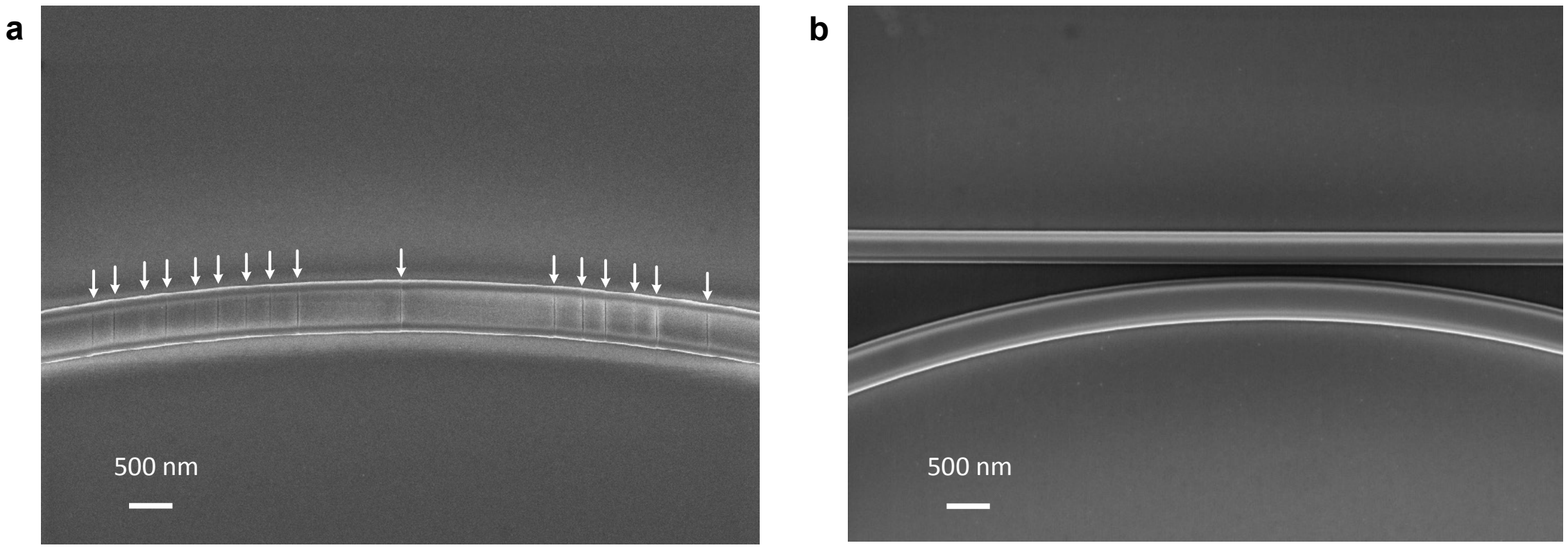

**Fig. 2 | Electron beam lithography (EBL) patterning comparison.** SEM pictures of developed curved pattern in e-beam resist HSQ with different EBL processes: **a,** Standard EBL process results in severe stitching (denoted by arrows) between fractured pattern segments. The stitching effect may be due to resist charging in EBL process[3]. **b,** The multi-pass EBL process mitigates the stitching effect and enables smooth curved waveguide patterning.

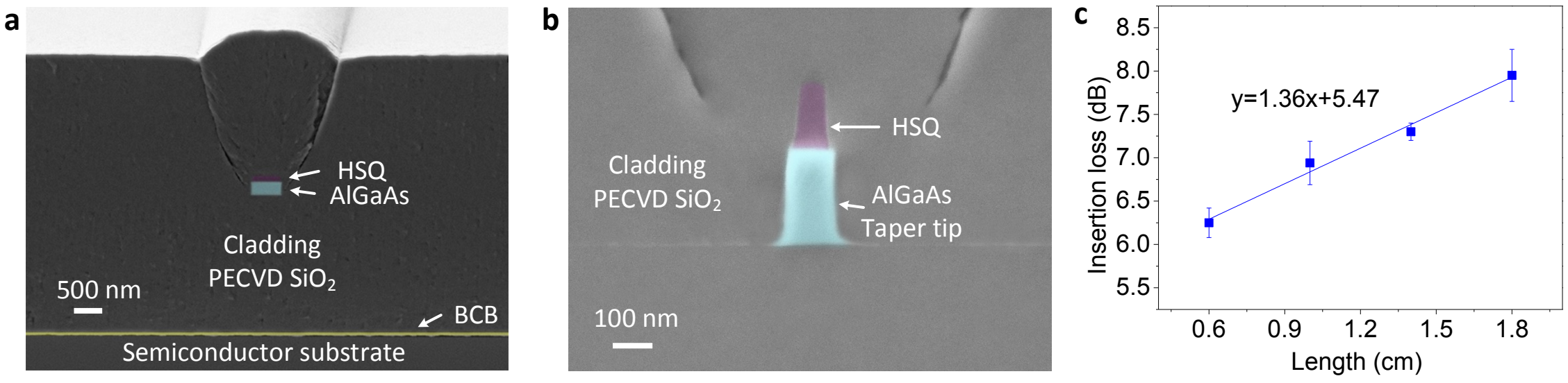

**Fig. 3 | Fabricated AlGaAsOI waveguide device and its loss measurment**. **a**, SEM cross-section picture of an AlGaAsOI nano-waveguide (denoted by the artificial blue colour) cladded in $SiO_2$. Benzocyclobutene (BCB) was used as bonding layer as denoted by the yellow colour. A strong light confinement can be achieved due to a small effective mode area[4] (0.16 $\mu m^2$) of the waveguide with cross-section dimension of 320 nm×630 nm. **b**, SEM picture of the nanotaper tip end at the sample facet. **c**, Measured insertion loss of AlGaAsOI waveguides with different lengths showing a linear loss of about 1.4 dB/cm. We measured three waveguides for each length. The waveguides used in the modified cut-back measurement have the same number of bends.

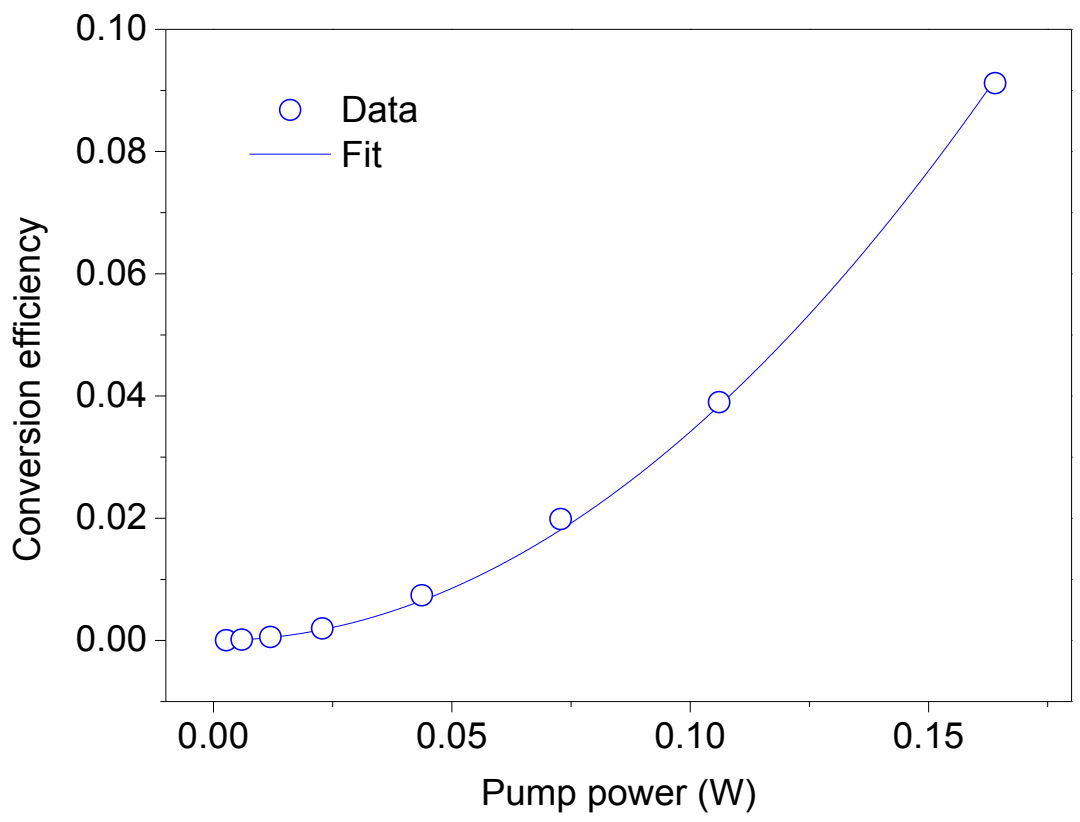

**Fig. 4 | Nonlinear coefficient estimation** by fitting the measured conversion efficiency versus coupled pump power for four-wave mixing in a 3-mm long AlGaAsOI waveguide with cross-section dimension of 320 nm×630 nm. An ultra-high effective nonlinearity of about 660 $W^{-1}m^{-1}$ was obtained and the derived material nonlinearity[5] of $Al_{0.17}Ga_{0.83}As$ is about $2.6\times10^{-17}$ $W/m^2$, which is similar to typical values previously reported[6-8].

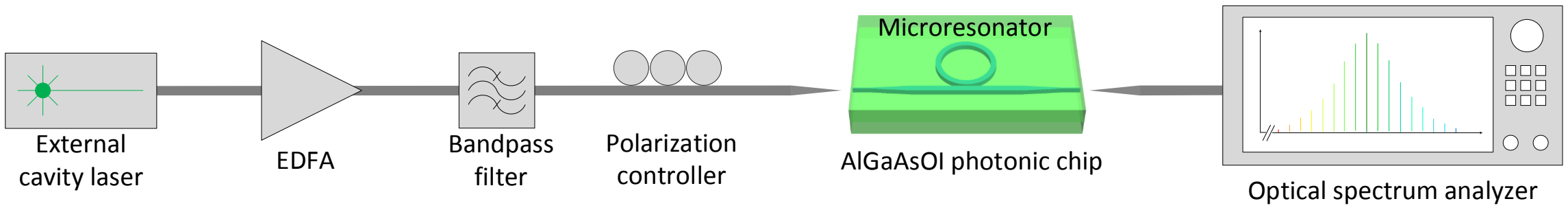

**Fig. 5 | Experimental set-up for frequency comb generation** based on the Kerr effect in a micro-resonator. A continuous wave pump laser (Ando AQ4321D), amplified by an erbium-doped fibre amplifier (EDFA, Amonics AEDFA), was passed through a tuneable band-pass filter (3-dB bandwidth: 16 nm) to minimize amplified spontaneous emission noise. Two lensed fibres were used for fibre-to-chip coupling. The output spectrum was measured with an optical spectrum analyser (Yokogawa AQ6375). The temperature of the sample substrate was kept at 22.5°C to ensure stable coupling. The input and output power were always monitored by a power meter (Hewlett Packard 8153A, not shown) during the experiment.

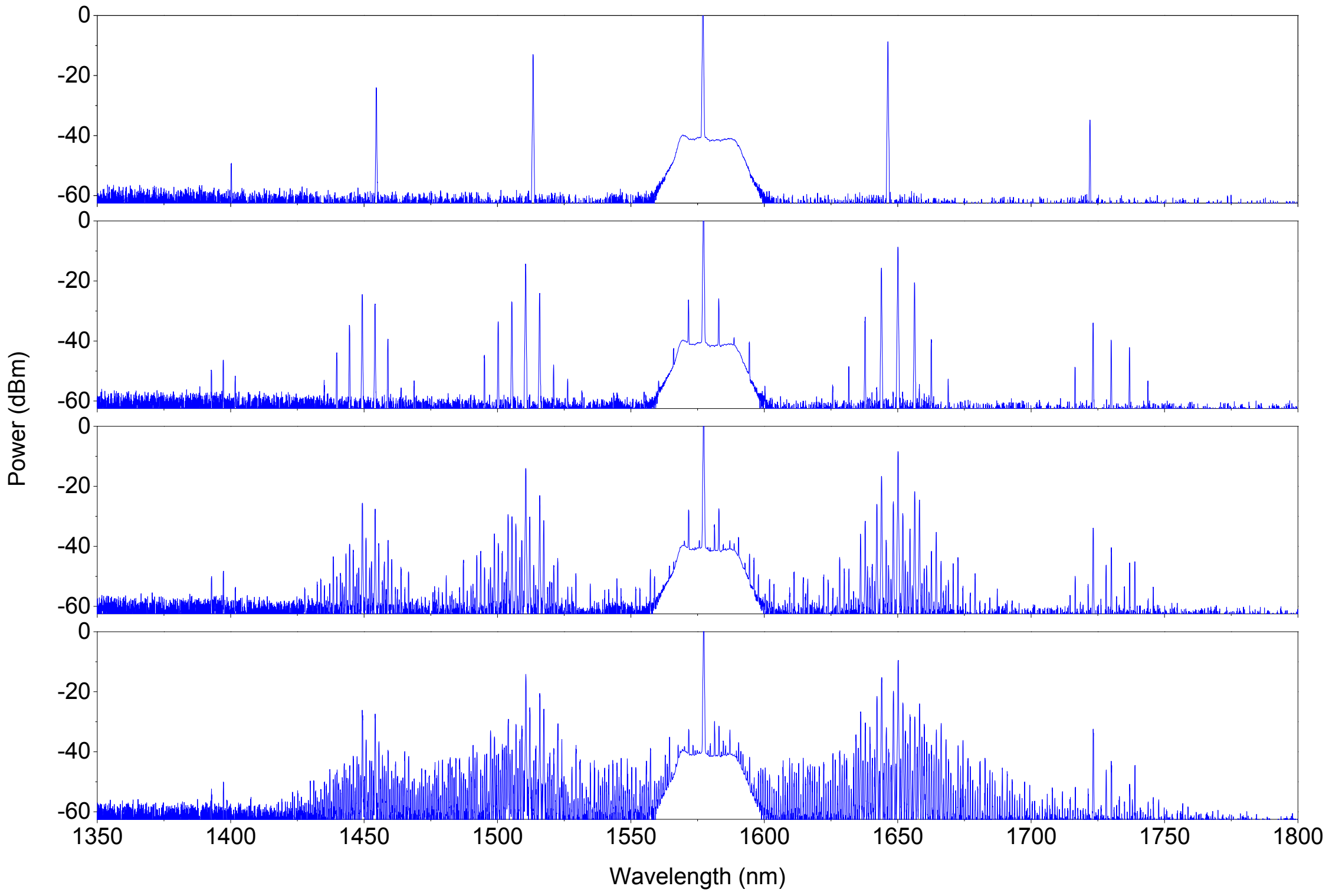

**Fig. 6 | Evolution of Kerr frequency comb spectrum** when the pump-resonance detuning is decreased (from the top spectrum to the bottom spectrum). Following primary sideband generation, sub comb lines were generated close to the primary sidebands and finally merged to form a frequency comb with the native line spacing (single-free spectral range), which shows a typical evolution as reported[9]. The coupled pump power was kept the same (42 mW in this case) during the pump wavelength tuning.

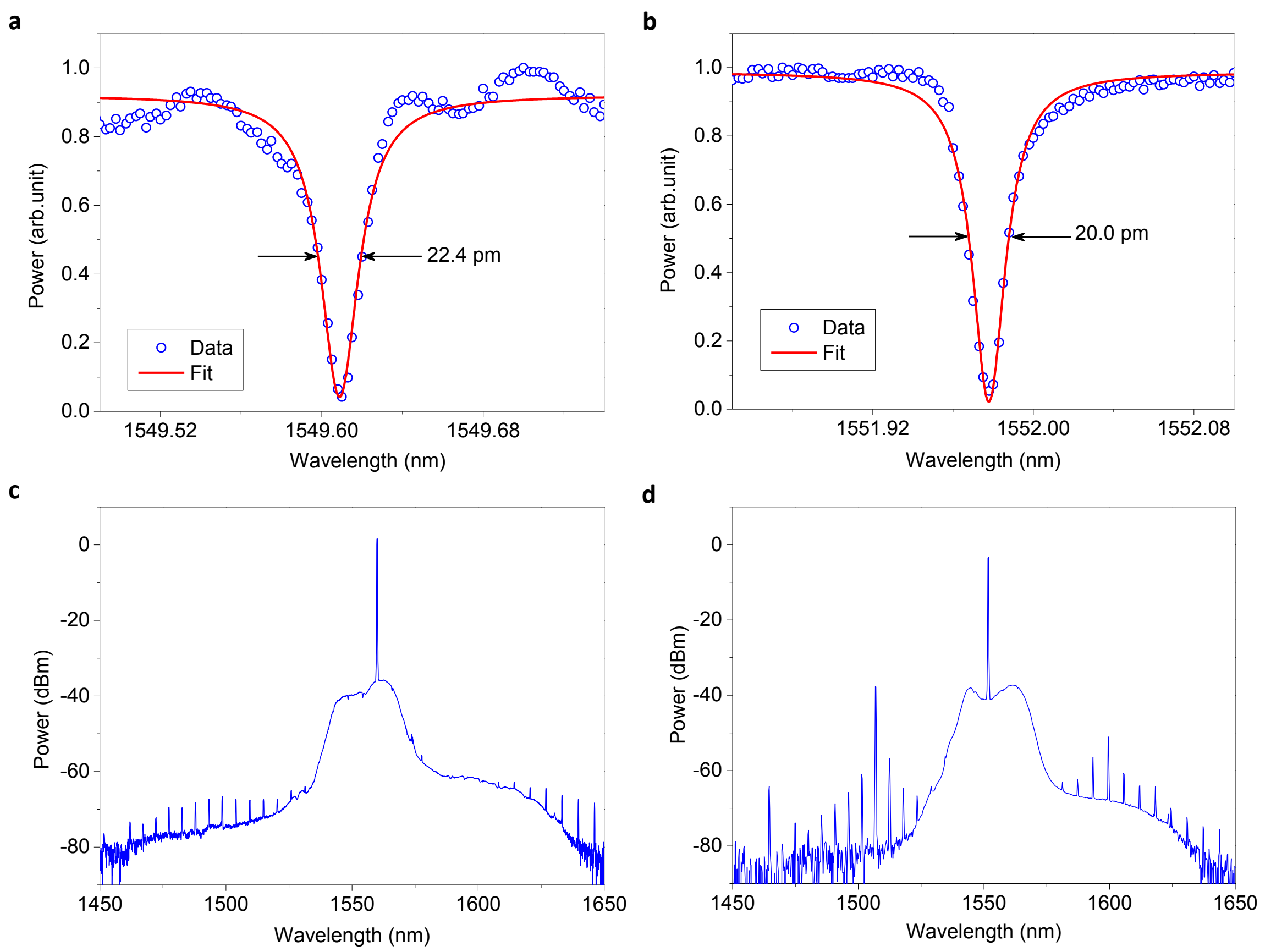

**Fig. 7 | Frequency comb generation performances of AlGaAsOI micro-resonators** with different aluminium content. **a,b,** Transmission spectra for resonances for the AlGaAsOI micro-resonator with 15% (**a**) and 17% (**b**) aluminium concentrations. **c,d,** Measured output spectra when the pump is fine-tuned to a resonance for the micro-resonator with 15% (**c**) and 17% (**d**) aluminium concentration. The coupled pump powers are 30 mW and 7 mW in **c** and **d**, respectively. As the bandgap[10] of $Al_{0.15}Ga_{0.85}As$ (1.61 eV) is not large enough to suppress two-photon absorption (TPA) efficiently[8], optical parametric oscillation (OPO) was not observed in the micro-resonator as shown in **c**. However, the nonlinear loss can be significantly reduced to a negligible level as the bandgap is increased[8]. Therefore, a slightly larger bandgap e.g. 1.64 eV ($Al_{0.17}Ga_{0.83}As$) results in a significant reduction of TPA and thus enables OPO in the micro-resonator with much lower pump power as shown in **d**.

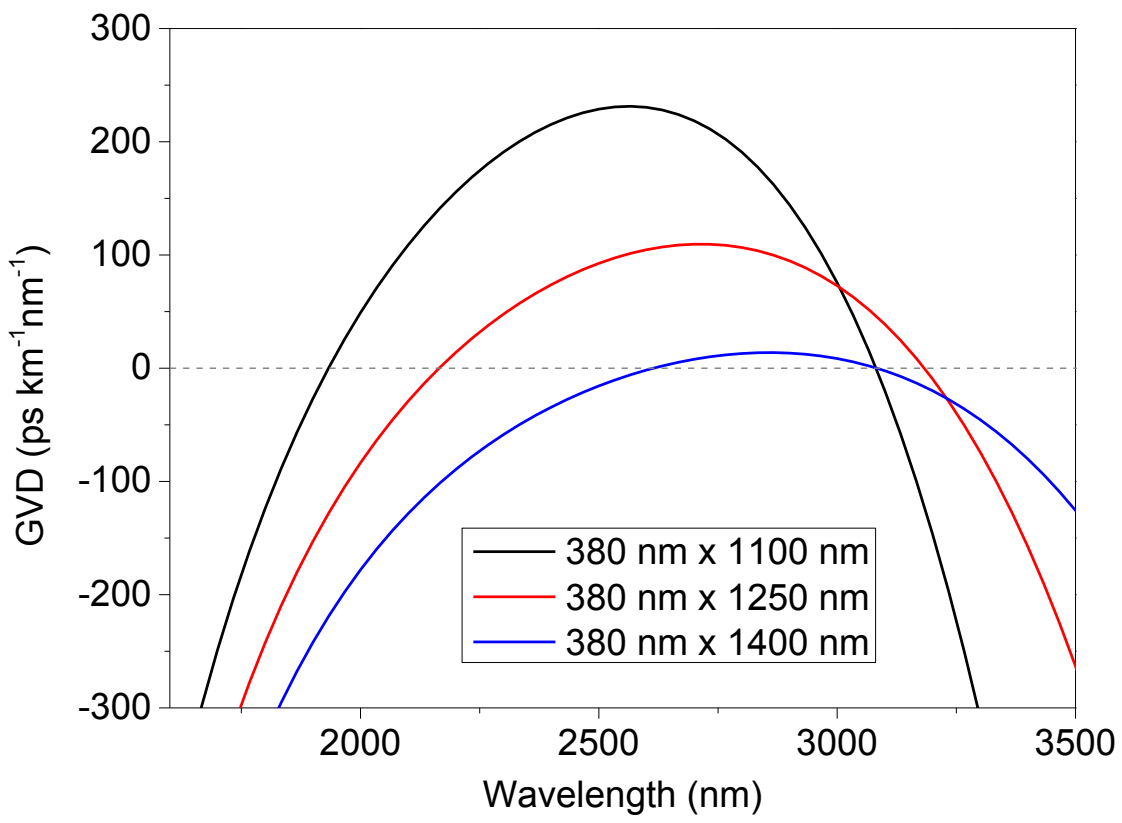

**Fig. 8 | Dispersion engineering of AlGaAsOI waveguide beyond 2 µm.** Anomalous dispersion can be obtained in mid-infrared wavelength range with larger dimension waveguides, which potentially allows mid-infrared frequency comb generation for applications such as molecular spectroscopy[11].